\documentclass[aps,prb,twocolumn,showpacs,showkeys,superscriptaddress]{revtex4}
\usepackage{graphicx}
\usepackage{dcolumn}
\usepackage{amssymb}

\begin{document}

\title{Phonon anomalies and charge dynamics in Fe$_{1-x}$Cu$_{x}$Cr$_{2}$S$_{4}$ single crystals}

\author{T.~Rudolf}
\affiliation{EP~V, Center for Electronic Correlation and
Magnetism, University of Augsburg, 86135~Augsburg, Germany}

\author{K.~Pucher}
\affiliation{EP~V, Center for Electronic Correlation and
Magnetism, University of Augsburg, 86135~Augsburg, Germany}

\author{F.~Mayr}
\affiliation{EP~V, Center for Electronic Correlation and
Magnetism, University of Augsburg, 86135~Augsburg, Germany}

\author{D.~Samusi}
\affiliation{Institute of Applied Physics, Academy of
Sciences of Moldova, MD 2028, Chisinau, Republic of Moldova}

\author{V.~Tsurkan}
\affiliation{EP~V, Center for Electronic Correlation and Magnetism,
University of Augsburg, 86135~Augsburg, Germany}
\affiliation{Institute of Applied Physics, Academy of Sciences of
Moldova, MD 2028, Chisinau, Republic of Moldova}

\author{R.~Tidecks}
\affiliation{Institute of Physics, University of Augsburg,
86135~Augsburg, Germany}

\author{J.~Deisenhofer}
\affiliation{EP~V, Center for Electronic Correlation and Magnetism,
University of Augsburg, 86135~Augsburg, Germany}

\author{A.~Loidl}
\affiliation{EP~V, Center for Electronic Correlation and
Magnetism, University of Augsburg, 86135~Augsburg, Germany}

\date{\today}

\begin{abstract}
A detailed investigation of phonon excitations and charge carrier
dynamics in single crystals of Fe$_{1-x}$Cu$_{x}$Cr$_{2}$S$_{4}$
($x= 0, 0.2, 0.4, 0.5$) has been performed by using infrared
spectroscopy. In FeCr$_{2}$S$_{4}$ the phonon eigenmodes are
strongly affected by the onset of magnetic order. Despite enhanced
screening effects, a continuous evolution of the phonon excitations
can be observed in the doped compounds with $x = 0.2$ (metallic) and
$x =0.4, 0.5$ (bad metals), but the effect of magnetic ordering on
the phonons is strongly reduced compared to $x=0$. The Drude-like
charge-carrier contribution to the optical conductivity in the doped
samples indicates that the colossal magneto-resistance effect
results from the suppression of spin-disorder scattering.
\end{abstract}




\maketitle

\section{Introduction}
The discovery of colossal magnetoresistance (CMR) in perovskite-type
manganites has attracted considerable attention.\cite{Kusters378,
Jirak374, Von360, Chahara371, Jin373} Double-exchange (DE)
mechanism, \cite{Zener365, Millis362} strong electron-phonon
coupling, \cite{Millis362} phase separation scenarios \cite{Mayr387}
or a Griffiths singularity~\cite{Salamon} were suggested to clarify
the origin of the CMR effect, but a conclusive microscopic model has
not yet been established. Ever since, the occurrence of CMR effects
has been reported for various other classes of materials, such as
pyrochlores,\cite{Shimakawa96} rare-earths based compounds like
GdI$_2$,\cite{Felser99} and ternary chalcogenide spinels
$A$Cr$_2$S$_4$.\cite{Ramirez255} These CMR materials have been
classified in terms of spin-disorder scattering and a universal
dependence of the magnetoresistence vs.~carrier density has been
suggested on theoretical grounds.\cite{Majumdar98,Gomez-Santos04}

Ramirez \emph{et al.}~drew attention to the spinel system
\mbox{Fe$_{1-x}$Cu$_x$Cr$_2$S$_4$} in 1997.\cite{Ramirez255} In
polycrystalline \mbox{FeCr$_2$S$_4$} with $T_C=170$~K, the CMR
effect reaches values comparable to those observed in perovskite
oxides. The substitution of Fe by Cu increases $T_C$ to
temperatures above room temperature, and the CMR effect remains
relatively strong ($\sim$7~\%).\cite{Ramirez255} In addition,
solid solutions of the ferrimagnetic semiconductor FeCr$_2$S$_4$
and the metallic ferromagnet CuCr$_2$S$_4$ show a number of
puzzling properties: From the very beginning, a controversial
discussion has been arising whether the Cu ions are mono- or
divalent for $x\geq
0.5$.\cite{Lotgering380,Goodenough110,Kurmaev379} For $x<0.5$ it
was established that only monovalent and hence diamagnetic
($d^{10}$) Cu exists in the mixed crystals.\cite{Tsurkan32}
Moreover, Fe$_{1-x}$Cu$_x$Cr$_2$S$_4$ shows two metal-to-insulator
transitions as a function of $x$, as the room-temperature
resistivity reveals two minima at $x=0.2$ and $x=1$ and
concomitantly the Seebeck coefficient changes sign two
times.\cite{Lotgering380,Haacke} Additionally, band-structure
calculations predicted that the Fe$_{1-x}$Cu$_x$Cr$_2$S$_4$ system
should exhibit a half-metallic
nature.\cite{Pickett01,Park213,Kurmaev379}

Recent experimental investigations of
\mbox{Fe$_{1-x}$Cu$_x$Cr$_2$S$_4$} single crystals indicated a
strong dependence of their magnetic and magneto-transport
properties on hydrostatic pressure suggesting a strong
magneto-elastic coupling.\cite{Tsurkan347, Fritsch395}
Measurements on the ac susceptibility in pure FeCr$_2$S$_4$
exhibited a cusp in the low-field magnetization and the onset of
magnetic irreversibilities at 60~K was explained by
domain-reorientation processes.\cite{Tsurkan346} Later on,
ultrasonic studies indicated an anomaly in the temperature
dependence of the shear modulus close to 60~K, and it was
suggested that the onset of orbital order induces a structural
distortion at this temperature.\cite{Maurer396} This result,
however, is hardly compatible with the observation that orbital
order is established in polycrystals close to 10~K, while an
orbital glass state is found in single crystals.\cite{Fichtl04}

Optical spectroscopy simultaneously probes the lattice and
electronic degrees of freedom and is, therefore, ideally suited to
investigate structural phase transitions and to clarify the
importance of electron-phonon coupling for the CMR
effect.\cite{Hartinger04} Earlier infrared (IR) studies in
polycrystalline \mbox{FeCr$_2$S$_4$} reported, in accordance with
the crystal-lattice symmetry $Fd3m$, the existence of four
IR-active phonons, which strongly depend on temperature near and
below $T_C$.\cite{Lutz375,Wakamura351} We performed measurements
of the optical properties of single crystals of
Fe$_{1-x}$Cu$_x$Cr$_2$S$_4$ ($x=0, 0.2, 0.4$ and 0.5) to shed
light on the interplay of structural and electronic properties in
these compounds. Since the optical properties of the samples with
$x=0.4$ and $x=0.5$ were found to be very similar, we only show
and discuss the corresponding data for $x=0.5$ in the following.

\section{Experimental details}
Single crystals of \mbox{Fe$_{1-x}$Cu$_x$Cr$_2$S$_4$} were grown
using a chemical transport-reaction method with chlorine as
transport agent and the ternary polycrystals as starting material.
Details of the sample preparation are described
elsewhere.\cite{Kurmaev379} No indication for the existence of
secondary phases was found by x-ray diffraction analysis of powdered
single crystals. X-ray single-crystal analysis confirmed the high
structural homogeneity of the samples. The composition and
homogeneity of the samples were examined by electron-probe
microanalysis. The samples were optically polished platelets with
dimensions of about $3\times5\times1$~mm$^3$. Structural, magnetic
and electrical transport data are given in
Ref.~\onlinecite{Fritsch395}.

Two Fourier-transform-infrared spectrometers with a full bandwidth
of 10 to 8000~cm$^{-1}$ ({\sc Bruker} IFS~113v) and 500 to
42000~cm$^{-1}$ ({\sc Bruker} IFS 66v/S) together with a $^4$He
cryostat ({\sc Oxford} Optistat) were used for measurements of the
optical reflectivity in the energy range from 70 to
30000~cm$^{-1}$ due to small sample dimensions and for
temperatures of \mbox{$5~\text{K}<T<300~\text{K}$}. In order to
investigate small fractions of the sample surface in the range of
0.1~mm$^2$ we utilized an IR microscope ({\sc Bruker} IRscope II),
which works in the far- (FIR) and mid-infrared (MIR) range.

\section{Experimental results and discussion}
\subsection{Phonon excitations}

\begin{figure}[t]
\includegraphics[angle=0,width=85mm,clip]{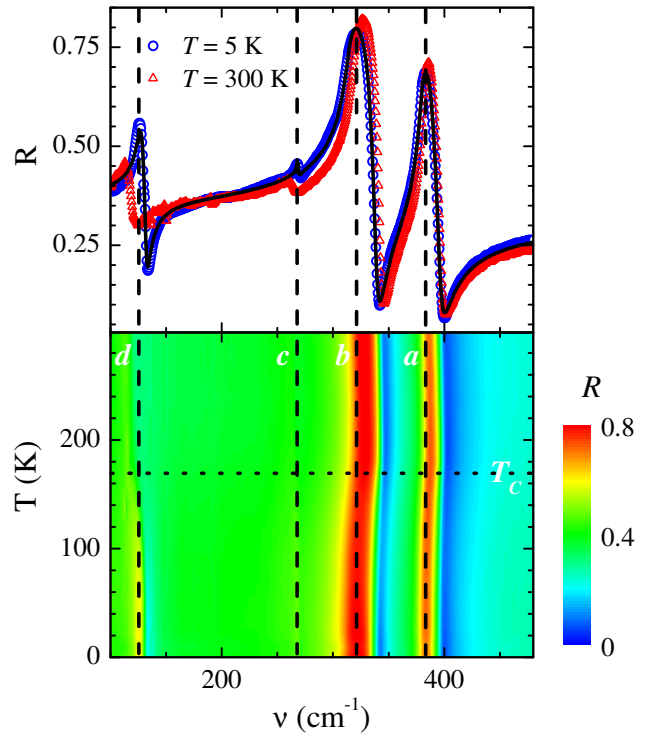}
\caption{(Color online) Upper panel: Reflectivity $R$ of
\mbox{FeCr$_2$S$_4$} vs.~wave number for $T=5$~K and 300~K. A fit of
the reflectivity for $T=5$~K is indicated by the solid line. Lower
panel: 2D-contour plot of the reflectivity $R$ vs.~$\nu$ and $T$
generated by interpolation of 17 spectra. The vertical lines are
highlighting the maxima of the IR-active-phonons in $R$ at 5~K.}
\label{fig1}
\end{figure}

Figure~\ref{fig1} shows the temperature dependence of the FIR
reflectivity $R$ vs.~wave number of pure \mbox{FeCr$_2$S$_4$}. In
the upper panel $R$ is plotted for 5 and 300~K. The four visible
phonon peaks are attributed to the four IR-active $F_{1u}$ modes
(symmetry group $Fd3m$, \#227).\cite{Lutz375} To analyze the
spectra, we used a 4-parameter fit assuming frequency-dependent
damping constants to account for the asymmetry of the phonon peaks.
This fitting procedure infers a splitting of the longitudinal and
transverse eigenfrequencies, $\omega_L$ and $\omega_T$, and the
corresponding damping constants, $\Gamma_L$ and $\Gamma_T$.
\cite{Berreman392} The resulting curves describe the measured
reflectivity down to 100~cm$^{-1}$ very well, without assuming an
additional contribution of free charge carriers. A representative
result of these fits is shown by the solid line in the upper panel
of Fig.~\ref{fig1} for $T=5$~K. The detailed temperature dependence
of the reflectivity is visualized in the two-dimensional (2D)
contour plot in the lower panel of Fig.~\ref{fig1}. To enable a
comparison of the phonon shift, the peak positions (maxima in $R$)
for $T=5$~K are indicated as vertical lines. Around $T_C=167$~K a
shift of the phonon frequencies can be observed, especially for the
mode $d$ close to 100~cm$^{-1}$. The intensity of this mode strongly
depends on temperature, too (see upper frame of Fig.~\ref{fig1}).

\begin{figure}[t]
\includegraphics[angle=0,width=85mm,clip]{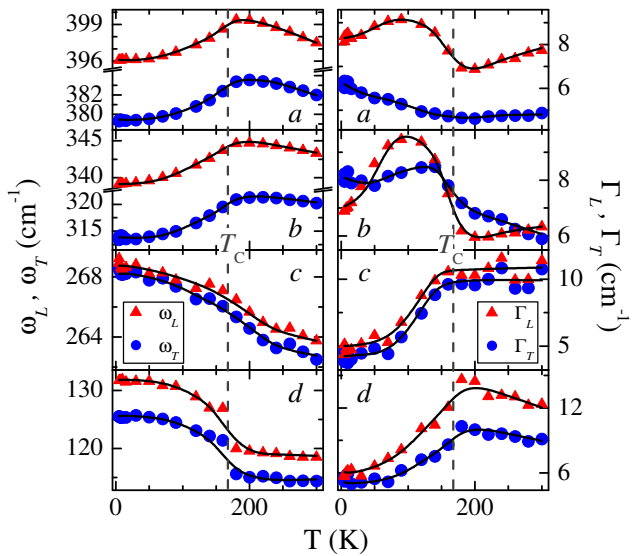}
\caption{(Color online) Temperature dependence of the longitudinal
(transverse) resonance frequencies $\omega_L$ ($\omega_T$) and
damping constants $\Gamma_L$ ($\Gamma_T$) obtained by a 4-parameter
fit for the four IR-active phonons in \mbox{FeCr$_2$S$_4$} as
described in the text. All solid lines are drawn to guide the eye.}
\label{fig2}
\end{figure}

The resonance frequencies $\omega_L$ and $\omega_T$ (left frames)
and the corresponding damping rates $\Gamma_L$ and $\Gamma_T$ (right
frames) are shown in Fig.~\ref{fig2} as a function of temperature.
Above the Curie temperature $T_C=167$~K, the resonance frequencies
$\omega_L$ and $\omega_T$ of all modes reveal a similar quasi-linear
increase with decreasing temperature, which can be fully ascribed to
anharmonic contributions to the lattice potential.\cite{Wakamura383}
In contrast to the rather usual behavior in the paramagnetic regime,
modes $a$ and $b$ soften for temperatures below $T_C$, while
$\omega_L$ and $\omega_T$ increase towards lower temperatures in the
case of modes $c$ and $d$. These anomalous changes of the
eigenfrequencies in the vicinity of $T_C$ suggest a correlation with
the onset of magnetic order. However, it has to be stated that the
size of the effect is different for the observed modes:
$\Delta\omega=[\omega(T=T_C)-\omega(T=0~$K$)]/\omega(T=0~$K$)$ is of
the order of $+3$~\% for the internal mode $d$, $\lesssim~+1$~\% for
the bending mode $b$, approximately $-1.5$~\% for the bending mode
$c$, and $-1$~\% for the stretching mode $a$. Longitudinal and
transverse eigenfrequencies behave rather similar.

The influence of magnetic order on phonons in magnetic
semiconductors has been proposed by Baltensperger and
Helman\cite{Baltensperger26} and Baltensperger\cite{Baltensperger27}
more than 30 years ago, and has recently been used by Sushkov
\textit{et al.} to describe the phonon spectra in
ZnCr$_2$O$_4$.\cite{Sushkov04} Based on a model calculation, where
superexchange interaction between the magnetic ions infers a
spin-phonon coupling, relative frequency shifts up to 10$^{-2}$ have
been predicted. The order of magnitude of this effect corresponds
nicely to the experimentally observed values in FeCr$_2$S$_4$ and,
therefore, Wakamura\cite{Wakamura351} considered this mechanism to
dominate the phonons' behavior for $T\lesssim T_C$. Subsequently,
Wakamura and coworkers\cite{Wakamura383,Wakamura} discussed the sign
of the relative frequency shift in terms of nearest-neighbor FM
exchange and next-nearest-neighbor AFM exchange for CdCr$_2$S$_4$,
which exhibits phonon modes with a similar temperature dependence as
FeCr$_2$S$_4$. Moreover, they could show that these anomalous
changes in the phonon frequencies are absent in non-magnetic
CdIn$_2$S$_4$, further corroborating their approach.\cite{Wakamura}
Thus, the positive shift of modes $a$ and $b$ would indicate that FM
exchange (Cr-S-Cr) dominates in accordance with a strong influence
of the (Cr-S) force constants on these modes, and, correspondingly,
the negative shift of modes $c$ and $d$ favors AFM exchange
(Cr-S-Cd-S-Cr) with a strong influence of the (Cd-S) force
constants. Note that a more rigorous theoretical treatment of
anharmonic spin-phonon and phonon-phonon interactions in cubic
spinels by Wesselinova and Apostolov\cite{Wesselinova96} confirms
the above interpretation. In \mbox{FeCr$_2$S$_4$} the interpretation
of the effect of magnetic ordering on the IR active phonon modes
becomes even more complicated, because there exist, besides FM
nearest-neighbor Cr-S-Cr bonds, additional exchange paths via AFM
Fe-S-Fe and Fe-S-Cr-S-Fe bonds. Nevertheless, the overall
temperature behavior of the phonon frequencies in FeCr$_2$S$_4$ is
similar to CdCr$_2$S$_4$ and may be well interpreted, accordingly.
Note, however, that a critical discussion of the above approach is
given by Bruesch and d'Ambrogio.\cite{Bruesch}

A straightforward interpretation of the temperature dependence of
the damping constants (right panel of Fig.~\ref{fig2}) is not
obvious at all. Again, considering only the anharmonicity of ionic
non-magnetic crystals, the damping is expected to show some residual
low-temperature value and a quasi-linear increase in the
high-temperature limit, just as observed for the longitudinal
damping constants of modes $a$ and $b$ for
$T>T_C$.\cite{Wakamura351} However, the temperature dependence of
$\Gamma_L$ and $\Gamma_T$ in general deviates from such a behavior:
In the case of mode $d$ both damping constants show a broad maximum
just above $T_C$ and a steep decrease towards lower temperature for
$T<T_C$. Mode $c$ follows a similar temperature dependence for
$T<T_C$, but the reduction of the damping constants is slightly
smaller, and in the paramagnetic regime $\Gamma_L$ and $\Gamma_T$
remain almost constant in contrast to the results of
Wakamura.\cite{Wakamura351} The behavior of modes $a$ and $b$ for
$T\lesssim T_C$  appears even more complex, but one can identify the
onset of enhancement damping close to $T_C=170$~K followed by broad
cusp-like maxima close to 100~K, except for $\omega_T$ of mode~$a$
that increases linearly with decreasing temperatures.

Wakamura\cite{Wakamura351} argues that the maxima of mode $d$ (and
$c$) are due to spin fluctuations of the Fe spins, in agreement with
the strong influence of the corresponding force constant on this
mode according to Bruesch and d'Ambrogio.~\cite{Bruesch}
Furthermore, long range spin order assumingly leads to the anomalous
changes of the damping constants for all modes below $T_C$. In
comparison to the temperature dependences of the damping constants
in \mbox{CdCr$_2$S$_4$}, one finds that modes $c$ and $d$ behave
similar to the case of \mbox{FeCr$_2$S$_4$}.\cite{Wakamura383} On
the other hand, modes $a$ and $b$ in \mbox{FeCr$_2$S$_4$} clearly
reveal a more complex behavior than in \mbox{CdCr$_2$S$_4$},
indicating a significant influence of the iron sublattice and the
additional effective exchange coupling between Fe-Fe and Fe-Cr ions
on these modes.

Additionally, we want to mention the large increase in intensity
(about 20~\%) for mode $d$ (close to 120~cm$^{-1}$) when cooling
from room temperature to 5~K (see Fig.~1). The intensity remains
almost constant above 200~K, while a linear increase with decreasing
temperature is observed below 200~K. At this temperature, maxima
appear in the temperature dependence of the damping constants,
suggesting a correlation of the two phenomena with regard to the
spin-fluctuation scenario discussed above.

When adopting the overall interpretation of the data in terms of
spin-phonon coupling, one has to consider, however, that e.g.~the
appearance of the cusps in the damping constants may be connected
to domain reorientation processes visible in the ac
susceptibility\cite{Tsurkan346} and anomalies detected by
ultrasonic investigations.\cite{Maurer396} Although the absence of
significant changes of the phonon frequencies contradicts the
scenario of a structural phase transition at 60~K driven by
orbital ordering as suggested in Ref.~\onlinecite{Maurer396}, it
becomes clear that the complex mechanisms dominating the damping
effects demand further theoretical studies to single out the
important contributions in detail.

\begin{figure}[t]
\includegraphics[angle=0,width=85mm,clip]{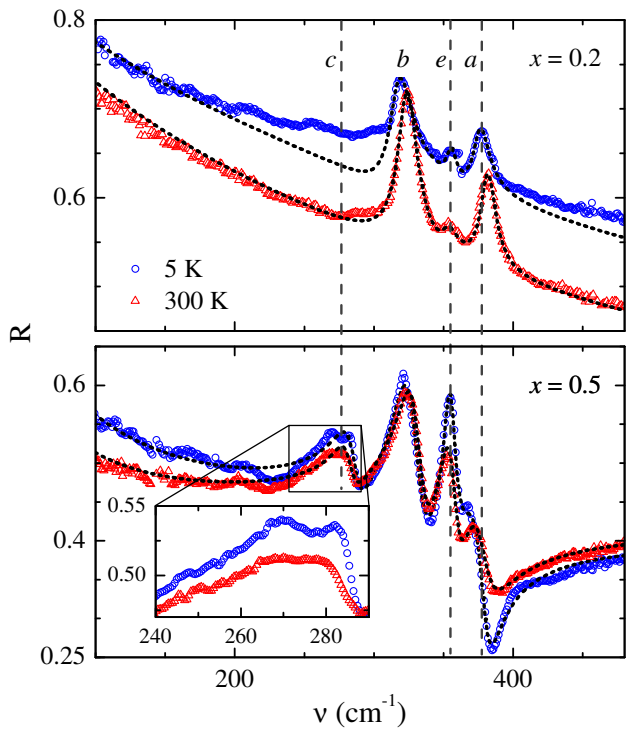}
\caption{(Color online) Reflectivity $R$ vs.~wave number for
\mbox{Fe$_{1-x}$Cu$_x$Cr$_2$S$_4$} with $x=0.2$ (upper panel) and
$x=0.5$ (lower panel) at $T=5$~K (open circles) and $T=300$~K (open
triangles). The dashed lines represent results of fits as described
in the text.} \label{fig3}
\end{figure}

Having discussed the phonon properties of pure FeCr$_2$S$_4$ we now
turn to the temperature dependence of the phonon modes for
Fe$_{1-x}$Cu$_x$Cr$_2$S$_4$. Figure~\ref{fig3} shows the FIR
reflectivity for $x=0.2$ (upper panel) and $x=0.5$ (lower panel) for
temperatures 5~K and 300~K each. The results for $x=0.4$ are very
similar to those obtained for $x=0.5$ and, hence, only the data for
$x=0.5$ is shown and discussed. The reflectivity of both samples,
$x=0.2$ and 0.5, shows a Drude-like contribution due to the presence
of free charge carriers, while FeCr$_2$S$_4$ can be described as an
insulator. The highest Drude-like conductivity is found for $x=0.2$
and the phonon modes are on the verge of being fully screened. For
both compounds the internal mode $d$ at $\sim120~$cm$^{-1}$ (see
Fig.~\ref{fig1} for the pure compound) can hardly be detected.
Focusing on the group of external modes, a new mode $e$ appears
close to $350~$cm$^{-1}$, while on increasing Cu concentration $x$
mode $a$ at $380~$cm$^{-1}$ becomes considerably reduced in
intensity. Without an accompanying lattice dynamical calculation one
cannot decide, if this new mode represents an impurity mode due to
the doping with Cu or a symmetry change. There are reports in
literature\cite{Lotgering380, Palmer384} claiming the reduction of
symmetry to $F\overline{4}3m$ because of the ordering of Fe and Cu
ions on the \textit{A} sublattice.
\begin{figure}[t]
\includegraphics[angle=0,width=85mm,clip]{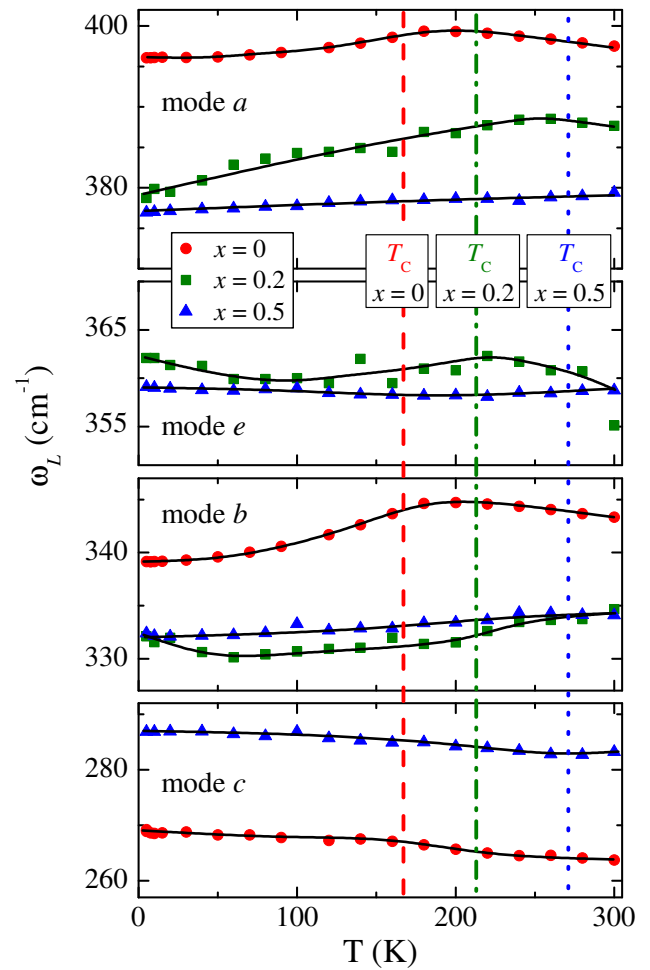}
\caption{(Color online) Temperature dependence of the longitudinal
eigenfrequencies $\omega_L$ of the IR-active phonon modes in
Fe$_{1-x}$Cu$_x$Cr$_2$S$_4$ for $x=0$~(circles), 0.2~(squares) and
0.5~(triangles). The dashed ($x=0$), dash-dotted ($x=0.2$) and
dotted lines ($x=0.5$) indicate the magnetic transition
temperatures. The solid lines are drawn to guide the
eye.}\label{fig4}
\end{figure}
In this case seven IR-active phonon modes are predicted. Assuming
that the peak close to 275~cm$^{-1}$ (mode $c$) is generated by two
single phonon modes (see inset in Fig.~3), five modes are visible,
only, while the internal mode $d$ close to $120~$cm$^{-1}$ remains
screened. However, our results rather point toward a continuous
evolution of the phonon modes on increasing $x$ and favor a
statistical \textit{A}-site distribution of Cu and Fe ions
throughout the lattice instead of a symmetry reduction resulting
from a superstructure due to an ordered \textit{A} sublattice.

We tried to fit the complete spectra taking into account the
reflectivity up to 10000~cm$^{-1}$, using a 4-parameter fit for the
phonon modes and a Lorentz oscillator for the mid-infrared
excitation at about $2500~$cm$^{-1}$ (see next section).
Representative results of these fits at low wave numbers are shown
in Fig.~\ref{fig3} as dashed lines. The temperature dependences of
the longitudinal modes $\omega_L$ as derived from these fits are
shown in Fig.~\ref{fig4} together with corresponding data for $x=0$.
The transverse eigenfrequencies behave rather similar (not shown).
Compared to the sample with $x=0$, the temperature dependence of the
damping constants in the doped compounds is weak and therefore not
shown here either. Regarding the sample with $x=0.2$ similar
anomalies as in pure FeCr$_2$S$_4$ can be seen in the vicinity of
$T_C$ for the observable modes $a,b$, and $e$. Obviously, the
temperature dependence of all phonon frequencies for $x=0.5$ is very
weak and no clear anomalies around $T_C$ are visible. Within the
experimental uncertainties one can detect a slight decrease of
$\omega_L$ towards lower temperatures except for mode $c$, which
behaves similarly to the case of FeCr$_2$S$_4$ (compare
Fig.~\ref{fig2}).

Keeping in mind the influence of spin fluctuations and spin-phonon
coupling on the phonon properties in FeCr$_2$S$_4$, Cu-doping seems
to reduce these features significantly. This observation is in
agreement with reduced spin-orbit coupling due to the substitution
of Jahn-Teller active Fe$^{2+}$ by non Jahn-Teller active Fe$^{3+}$.
Therefore, for $x=0.5$ only Fe$^{3+}$ with a half-filled $d$-shell
is present in the system\cite{Lang00,Kurmaev379} and the system
becomes almost magnetically isotropic as it was confirmed by
ferromagnetic resonance experiments.\cite{Fritsch395}

\subsection{Dynamic conductivity and electronic excitations}

\begin{figure}[t]
\includegraphics[angle=0,width=85mm,clip]{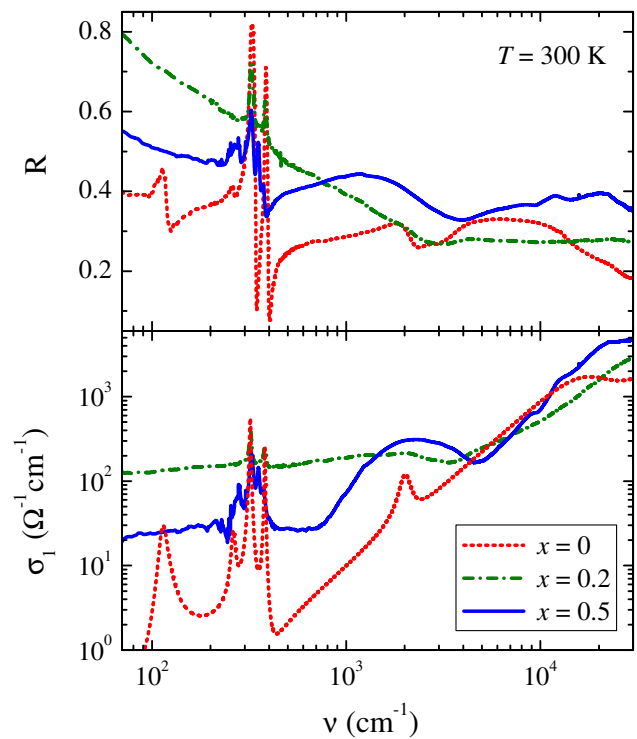}
\caption{(Color online) Upper panel: Semi-logarithmic plot of the
room-temperature reflectivity vs.~wave number in
Fe$_{1-x}$Cu$_x$Cr$_2$S$_4$ for Cu concentrations $x=0$, 0.2 and
0.5. Lower panel: Double-logarithmic plot of the real part of the
dynamic conductivity $\sigma_1$ as derived from the reflectivity
spectra.} \label{fig5}
\end{figure}

When the reflectivities of the doped compounds with Cu
concentrations $x=0.2$ and 0.5 (Fig.~\ref{fig3}) are compared with
that of pure FeCr$_2$S$_4$ it becomes clear that contributions from
free charge carriers have to be taken into consideration. The
metallic-like behavior is most significant for $x=0.2$, but it
becomes reduced again on further doping. For a consistent
description of the Drude-type behavior of the doped compounds, it is
important to measure the reflectivity spectra to higher energies.
The room-temperature reflectivities of Fe$_{1-x}$Cu$_x$Cr$_2$S$_4$
for $x=0.2$ and 0.5 are plotted in the upper panel of
Fig.~\ref{fig5} up to 3$\times$10$^4$~cm$^{-1}$, corresponding to
almost 4~eV, and are compared to the reflectivity of insulating
FeCr$_2$S$_4$. For the Kramers-Kronig analysis of the smoothed
reflectivity data we used a low-frequency Hagen-Rubens extrapolation
and a high-frequency extrapolation with a $\nu^{-0.5}$ power law up
to 10$^6$~cm$^{-1}$ and a subsequent $\nu^{-4}$ high-frequency tail.
The resulting dynamic conductivities $\sigma(\nu)$ are shown in the
lower panel of Fig.~\ref{fig5}. We carefully checked the
high-frequency extrapolation, also trying smoother extrapolations,
but found that the results are not influenced in the relevant energy
range below 20000~cm$^{-1}$. The use of a Hagen-Rubens extrapolation
is justified by the fact that we have the complete information on
the absolute values of the dc conductivities and the corresponding
temperature dependences for all compounds, although we are aware of
the additional uncertainties originating from the Hagens-Rubens
extrapolation, specifically for the sample with $x=0.5$. However,
the best fits of the reflectivity at room temperature, even in the
limited spectral range, yielded dc conductivities of
150~($\Omega$cm)$^{-1}$ for $x = 0.2$ and 35~($\Omega$cm)$^{-1}$ for
$x = 0.5$, close to the dc values derived from the 4-probe
measurements on single crystals by Fritsch \textit{et
al.}~\cite{Fritsch395}

For $x=0$ a weak but well defined electronic transition is observed
close to $2000~$cm$^{-1}$ and a further transition appears close to
20000~cm$^{-1}~(\approx 2.5$~eV). On substituting iron by copper,
metallic behavior shows up and for
\mbox{Fe$_{0.8}$Cu$_{0.2}$Cr$_2$S$_4$} the dc conductivity is of the
order 150~($\Omega$cm)$^{-1}$. The transition at $2000~$cm$^{-1}$,
becomes almost fully suppressed for $x=0.2$. Obviously, the
$d$-electrons become strongly delocalized. It is generally accepted
that in an ionic picture monovalent Cu is substituted inducing
trivalent Fe. Our results suggest that the system behaves as if
holes are doped into an insulator driving the compound into a
metallic regime. Unexpectedly, a broad peak appears again close to
2500~cm$^{-1}$ for $x=0.5$.

The observed doping dependence of the conductivity spectra as
documented in Fig.~\ref{fig5} can be compared with band-structure
calculations of these compounds.~\cite{Lang00,Park213,Kurmaev379}
Local spin-density approximation (LSDA) band-structure calculations
predict a half-metallic ground state of \mbox{FeCr$_2$S$_4$}, with a
partly filled $e$ band at the Fermi level. Correlation effects via
LSDA+\textit{U} yield a splitting of the Fe $e$ band into a lower
and upper Hubbard band characterizing FeCr$_2$S$_4$ as a
Mott-Hubbard insulator.\cite{Park213} The splitting of the $e$ band
is of the order of about 0.5~eV, and, hence, the peak close to
2000~cm$^{-1}$ may be interpreted as a transition between the lower
and upper Hubbard band. Accordingly, the high-energy excitation can
be attributed to a Cr(3\textit{d}) to Fe(3\textit{d}) transition.

Using an ionic picture with localized Fe $d$ states, alternatively,
the transition at 2000~cm$^{-1}$ may correspond to a transition
between the lower $e$ doublet and the $t_2$ triplet of the Fe
$d$-states split in a tetrahedral crystal-field. The expected
crystal-field splitting for Fe$^{2+}$ located in the tetrahedral
site of the spinel structure is rather weak\cite{Abragam70} and a
splitting of the order $2000-3000$~cm$^{-1}$ seems reasonable.
Further support for this interpretation comes from the observation
of crystal-field transitions as measured for diluted Fe$^{2+}$ in
CdIn$_2$S$_4$. Here a crystal-field splitting of approximately
2500~cm$^{-1}$ has been reported by Wittekoek \emph{et
al.}\cite{Wittekoek397}

The appearance of the broad excitation for $x=0.5$ in the
mid-infrared region at about 2500~cm$^{-1}$, however, cannot be
explained easily. In an ionic picture only trivalent iron and
monovalent copper are expected for
$x=0.5$,\cite{Lotgering380,Goodenough110} and recent x-ray
photoelectron spectroscopy~\cite{Tsurkan32} strongly favors the
existence of only monovalent Cu for $x=0.5$. Therefore, one can
exclude the possibility that the broad excitation may be attributed
to Fe$^{2+}$ similarly to the well-defined electronic excitation for
$x=0$. Nevertheless, it has been concluded from M\"{o}ssbauer
experiments in Fe$_{1-x}$Cu$_x$Cr$_2$S$_4$, that an ionic picture is
not applicable at all.~\cite{Haacke2} For $x=0.3$ and $T<T_C$ the
complicated M\"{o}ssbauer spectra indicate two different Fe sites
corresponding to Fe$^{2+}$ and Fe$^{3+}$, while for $T>T_C$ a single
line pointed towards a fast electron exchange between these two
sites. For $x=0.5$ the line pattern for $T>T_C$ evidenced the
existence of Fe$^{3+}$ and a strong delocalization of the Cu
$d$-derived electrons. Hence, further studies beyond the scope of
this paper are needed to clarify the nature of this mid-infrared
excitation.

In the following we will discuss the optical conductivity results in
the low frequency range in comparison with the dc conductivity data
reported in Ref.~\onlinecite{Fritsch395}. The room-temperature
spectra for the concentrations $x = 0.2$ and 0.5, shown in Fig.~5
have been used to estimate the Drude-like conductivity. For all
temperatures, the spectra could satisfactorily be described using a
plasma frequency $\omega_p = 12000$~cm$^{-1}$ and a dielectric
constant $\epsilon_\infty = 10.6$ for $x = 0.2$, which is close to
the value $\epsilon_\infty = 11.5$ for $x=0$. For $x = 0.5$ we used
$\omega_p = 5000~$cm$^{-1}$ and an enhanced dielectric constant
$\epsilon_\infty = 15.5$. The enhanced $\epsilon_\infty$ indicates
strong changes in the electronic excitation spectrum at higher
frequencies, but due to the complexity of the spectrum in this
energy region there is also a larger uncertainty in
$\epsilon_\infty$ for $x = 0.5$. The decrease of the plasma
frequency by a factor of 2.4 can be explained by a decrease of the
charge carrier density, as Fe$_{1-x}$Cu$_x$Cr$_2$S$_4$ approaches a
metal-to-insulator transition close to $x=0.5$. With these values,
the conductivity below 500~cm$^{-1}$ could reasonably be fitted for
all temperatures as indicated by the dashed lines in Fig.~3 for the
spectra at 5~K and 300~K.

\begin{figure}[t]
\includegraphics[angle=0,width=85mm,clip]{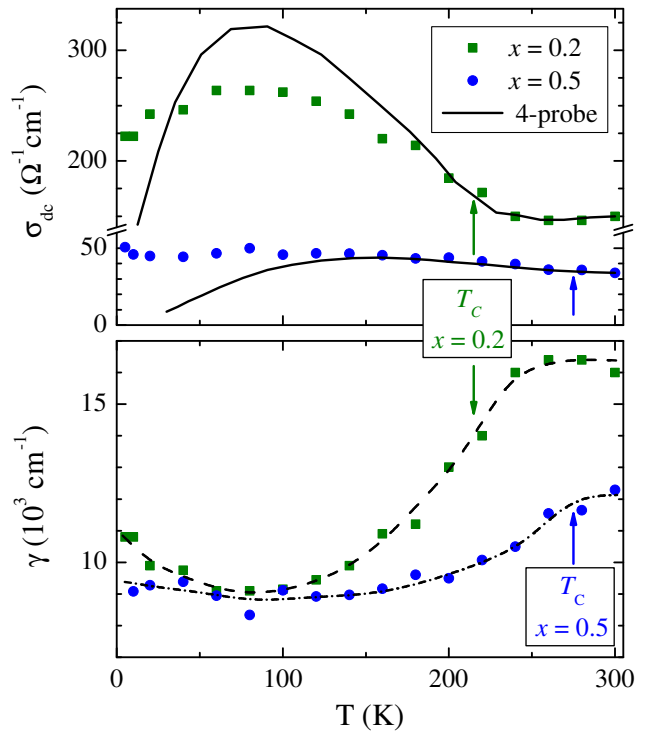}
\caption{(Color online) Upper panel: Temperature dependence of the
dc conductivity of Fe$_{1-x}$Cu$_x$Cr$_2$S$_4$ as determined from
fits to the reflectivity (see text). The dc conductivities as
observed from 4-probe measurements\cite{Fritsch395} are indicated by
solid lines and were scaled to the room temperature optical values.
Lower panel: Temperature dependence of the Drude-like relaxation
rates. Ferrimagnetic ordering temperatures are indicated by arrows.
The dashed and dash-dotted lines are drawn to guide the eye.}
\label{fig6}
\end{figure}

The resulting temperature dependences of the dc conductivity (upper
panel) and relaxation rates $\gamma\propto\tau^{-1}$ (lower panel)
are shown in Fig.~\ref{fig6}. The dc conductivities as derived from
4-probe measurements\cite{Fritsch395} are indicated by solid lines.
The dc conductivities were scaled at room temperature, utilizing a
factor of 1.6 for $x=0.2$ and a factor of 1.05 for $x=0.5$. Above
100~K the 4-probe dc results and the dc values as derived from the
optical measurements follow a similar temperature dependence.
However, at low temperatures the dc measurements are dominated by
localization effects, which appear much weaker in the high-frequency
($>100~$cm$^{-1}$) derived optical data. That localization effects
are most significant in the low-frequency ("dc") transport
measurements becomes clear from the fact that in doped
semiconductors the conductivity below the FIR regime increases
almost linearly with frequency.~\cite{Lunkenheimer} In the sample
with $x=0.5$, which exhibits the lower conductivity, localization
effects dominate already at higher temperatures. This may be
attributed to a significant decrease of the charge-carrier density
and concomitant increase of disorder due to the statistical
distribution of the Cu ions in the lattice,\cite{Lang00} further
discarding the possibility of $A$-site order of Fe and Cu for
Fe$_{1-x}$Cu$_x$Cr$_2$S$_4$.

Finally, we want to draw attention to the temperature dependence
of the relaxation rate $\gamma$ (lower panel of Fig.~6). In the
magnetically ordered state below $T_C$, the relaxation rates
become significantly reduced, e.g.~the reduction amounts to almost
50\% for $x=0.2$. We recall that the plasma frequency has been
kept constant for each compound as a function of temperature. This
indicates that the increase of the conductivity just below the
magnetic ordering temperature results from the freezing-out of
disorder scattering and not from a change of the carrier density
via band-structure changes at the onset of ferrimagnetic order.
Taking into account the classification of chalcogenide spinels
$A$Cr$_2$S$_4$ as systems where CMR originates from spin-disorder
scattering,\cite{Majumdar98} the observed reduction of the
relaxation rate below $T_C$ has to be regarded as direct evidence
of such a scenario: In external fields the onset of ferrimagnetic
order shifts to higher temperatures. Concomitantly, a reduction of
the scattering rate and the anomalous increase of the conductivity
arise. As a consequence, maximal CMR effects will show up just
below $T_C$ as a function of an external magnetic field. A similar
scenario has been reported for GdI$_2$, where the magnetic and
magneto-transport properties have been described successfully in
terms of spin-fluctuations and their suppression by external
magnetic fields in the vicinity of
$T_C$.\cite{Eremin01,Deisenhofer04} We would like to point out,
that at low temperatures the relaxation rates for $x=0.2$ and
$x=0.5$ are of the same order of magnitude $\sim 10^4$cm$^{-1}$,
indicating a similar level of disorder for the Cu doped compounds.

\section{Summary}

In summary, we investigated the optical properties of
Fe$_{1-x}$Cu$_x$Cr$_2$S$_4$ single crystals for Cu concentrations
$x=0$, 0.2, 0.4 and 0.5. Phonon excitations and dynamic
conductivity for $x=0.4$ are very similar to the results for
Fe$_{0.5}$Cu$_{0.5}$Cr$_2$S$_4$ and were not discussed separately.
The phonon excitations were measured as a function of temperature
between 5~K and room temperature. Pure FeCr$_2$S$_4$ shows clear
anomalies in the eigenfrequencies at the transition from the
paramagnetic to the ferromagnetic state, which can be explained by
spin-phonon coupling. Concerning the complex behavior of the
damping constants, spin fluctuations in the vicinity of $T_C$ may
describe many of the anomalous changes, but further theoretical
studies are necessary to corroborate this interpretation. The
influence of magnetic order on the eigenmodes is reduced with
increasing $x$, and the appearance of a new phonon mode close to
350~cm$^{-1}$ is attributed to an impurity mode rather than to a
symmetry reduction due to A-site order.

Morover, the charge dynamics of Fe$_{1-x}$Cu$_x$Cr$_2$S$_4$ were
investigated. FeCr$_2$S$_4$ is an insulator, but becomes metallic
when slightly doped with Cu. The conductivity of the free charge
carriers can be described by a normal Drude-type behavior. The dc
conductivity for $x=0.2$ is enhanced by a factor of four in
comparison to $x=0.5$. The temperature dependence of the optically
derived dc conductivity for both doped compounds is is in good
agreement with resistivity measurements, but localization effects
at lowest temperatures appear weaker in the optical measurements.
The corresponding behavior of the scattering rate, which shows a
strong decrease below the ferrimagnetic phase transition,
evidences the freezing-out of disorder scattering below $T_C$. In
accordance with the proposed classification of the ternary
chalcogenide spinels as spin-disorder magnetoresistive materials,
the reduction of the relaxation rate corroborates such a scenario
and makes clear that spin-disorder has to be considered a
necessary ingredient towards a theoretical description of this
fascinating class of materials.

\begin{acknowledgments}
It is a pleasure to thank H.-A.~Krug von Nidda, J.~Hemberger, and
Ch.~Hartinger for fruitful discussions. This work was partly
supported by the DFG via the Sonderforschungsbereich 484 (Augsburg),
by the BMBF/VDI via the Contract No.~\mbox{EKM/13N6917/0}, by the
U.S. Civilian Research \& Development Foundation (CRDF) and by the
Moldavian Research \& Development Association (MRDA) via Grant No.
MP2-3047.
\end{acknowledgments}


\end{document}